# Post deposition interfacial Néel temperature tuning in magnetoelectric B:Cr$_2$O$_3$

Ather Mahmood[1], Jamie Weaver[2], Syed Qamar Abbas Shah[1], Will Echtenkamp[1,3], Jeffrey W. Lynn[4], Peter A. Dowben[1], Christian Binek[1,*]

[1]Department of Physics & Astronomy and the Nebraska Center for Materials and Nanoscience, University of Nebraska-Lincoln, Lincoln, NE 68588-0299, USA

[2]Material Measurement Laboratory, National Institute of Standards and Technology, Gaithersburg, Maryland, USA 20899

[3]Department of Electrical & Computer Engineering, University of Minnesota, Minneapolis, MN 55455, USA

[4]NIST Center for Neutron Research, National Institute of Standards and Technology, Gaithersburg, MD 20899, USA

*To whom correspondence should be addressed. E-mail: cbinek@unl.edu

Boron (B) alloying transforms the magnetoelectric antiferromagnet Cr$_2$O$_3$ into a multifunctional single-phase material which enables electric field driven π/2 rotation of the Néel vector. Nonvolatile, voltage-controlled Néel vector rotation is a much-desired material property in the context of antiferromagnetic spintronics enabling ultra-low power, ultra-fast, nonvolatile memory, and logic device applications. Néel vector rotation is detected with the help of heavy metal (Pt) Hall-bars in proximity of pulsed laser deposited B:Cr$_2$O$_3$ films. To facilitate operation of B:Cr$_2$O$_3$-based devices in CMOS environments, the Néel temperature, $T_N$, of the functional film must be tunable to values significantly above room temperature. Cold neutron depth profiling and x-ray photoemission spectroscopy depth profiling reveal thermally activated B-accumulation at the B:Cr$_2$O$_3$/ vacuum interface in thin films deposited on Al$_2$O$_3$ substrates. We attribute the B-enrichment to surface segregation. Magnetotransport data confirm B-accumulation at the interface within a layer of about 50 nm thick where the device properties reside. Here $T_N$ enhances from 334 K prior to annealing, to 477 K after annealing for several hours. Scaling analysis determines $T_N$ as a function of the annealing temperature. Stability of post-annealing device properties is evident from reproducible Néel vector rotation at 370 K performed over the course of weeks.

# I. Introduction

Magnetic field and magnetization are axial or pseudovectors.[1] Manipulating the axial magnetization vector solely with the help of a polar electric field vector is a key challenge in the field of spintronics. However, electric field control of magnetization in the absence of dissipative electric currents is essential for energy efficient devices with applications in information technology. Of particular interest is energy-efficient, ultra-fast and non-volatile switching at temperatures which allow operation in a CMOS environment. Antiferromagnets, which avoid the slow (ns) magnetization reversal of ferromagnets, are candidates for ultra-fast spintronics. Voltage-controlled antiferromagnetic (AFM) spintronics promise energy efficiency due to the absence of dissipative electric currents required in spin transfer torque or spin orbit torques alternatives. Recently, boron (B) doping of the prototypical magnetoelectric antiferromagnet $Cr_2O_3$ revealed new functionalities beyond the magnetoelectric effect of undoped $Cr_2O_3$ (chromia), *i.e.*, induced magnetization (polarization) in the presence of an electric (magnetic) field. Pure chromia is the archetypical magnetoelectric antiferromagnetic material.[2-4] Below its antiferromagnetic ordering temperature, $T_N$, spatial and time inversion symmetry are broken while the combined operation of parity and time inversion leave the system invariant. Under these conditions, the linear magnetoelectric effect becomes symmetry allowed. A diagonal magnetoelectric susceptibility emerges with distinct magnetoelectric response parallel and perpendicular to the three-fold symmetry axis of the system.[3] The new functionalities associated with B-doping include an increase of the Néel temperature in as-deposited films from $T_N$= 307 K in $Cr_2O_3$ to about $T_N$= 400 K in B:$Cr_2O_3$ at about 3% of anion substitution by boron.[5-7] More surprisingly, in B:$Cr_2O_3$ a different type of magnetic response with an applied electric ($E$) field emerges. It enables isothermal voltage-controlled rotation of the Néel vector by $\pi/2$ in the absence of an applied magnetic field.[8] 180 degree switching of the Néel vector has been achieved in chromia when utilizing its magnetoelectric effect to lift the degeneracy between the two 180-degree domain states. In pure $Cr_2O_3$ switching requires simultaneous application of an $E$-field (polar vector) and magnetic ($H$) field (axial vector) along the crystallographic $c$-axis. The pseudoscalar product, $EH$, has to overcome a critical threshold $(EH)_c$ to initiate switching.[9-15]

Because voltage controlled Néel vector rotation in B:$Cr_2O_3$ takes place in $H$=0, the magnetoelectric effect, although still present in B:$Cr_2O_3$, is not causing the spin rotation. While details of the rotation mechanism are still under investigation, there is strong evidence from piezo force and magnetic force microscopy that B-doping breaks local spatial inversion symmetry. The local symmetry breaking creates polar nanoregions, which, in the presence of a homogenous $E$-field, give rise to uniform polarization. Piezo response accompanies the polarization, which strains the sample. Magnetoelastic coupling changes the anisotropy in response to the strain, which gives rise to non-volatile reorientation of the easy axis and reorientation of the Néel vector from in-plane to out-of-plane orientation and back.[16] Magnetotransport measurements are sensitive to the orientation of the Néel vector. A heavy metal (Pt) Hall-bar in proximity of B:$Cr_2O_3$ films enables detections of a transverse Hall voltage, $V_{xy}$, in response to a current density $j$ applied in the $x$-direction of the Hall bar where $z$ points normal to the plane. The voltage signal $V_{xy}$ can be used as a proxy for the boundary magnetization of the antiferromagnet. Boundary magnetization is an equilibrium property of insulating magnetoelectric antiferromagnets and is known from rigorous symmetry considerations[17] and the concept of magnetoelectric multipolization.[18] It describes a roughness insensitive surface magnetization, which intimately couples to the orientation of the Néel vector. Magnetotransport measurements are sensitive to the orientation of the boundary magnetization and thereby the orientation of the Néel vector. Experiments suggest that spin Hall magnetoresistance associated with interfacial spin scattering provides the dominant contribution to $V_{xy}$.[19-21] Alternative mechanisms include proximity effects and interfacial chiral spin structures.[22] Recent Hall measurements in large applied magnetic fields add to the notion that spin Hall magnetoresistance is not the sole contributor to the Hall signal.[23] Nevertheless, its use as proxy for orientation and temperature dependence of boundary magnetization and the associated AFM order parameter are well established.[11-13]

In this work, we show that the operation temperature for reliable, voltage controlled Néel vector toggling, can be thermally tuned within a temperature range as wide as $\Delta T \approx 200$ K via a post-deposition annealing protocol. First, we discuss data from cold Neutron Depth Profiling (cNDP) and x-ray photoemission spectroscopy (XPS) depth profiling which reveal that annealing transforms the B-concentration depth profile from its as-deposited virtually uniform profile into a peaked concentration profile with its maximum near the surface of the B:$Cr_2O_3$ film. These data are accompanied by magnetotransport measurements in B:$Cr_2O_3$/Pt Hall-devices. In agreement with the evolution of the concentration profile measured via cNDP and XPS, the transport data imply that the Néel temperature at the surface increases in response to annealing. We determine the dependence of the Néel temperature on the annealing temperature via a scaling analysis of the $V_{xy}$ versus $T$ data measured after various annealing protocols.

## 2. Sample preparation

Pulsed Laser Deposition (PLD) in ultra-high vacuum with a base pressure of $\approx 5 \times 10^{-9}$ Torr ($\approx 7 \times 10^{-7}$ Pa) is used to grow (0001)-oriented films of the sesquioxides $V_2O_3$ and subsequently, B-doped chromia. The $V_2O_3$ film, which is grown on the *c*-plane of a sapphire single crystalline substrate, serves as bottom electrode in a gated Hall bar structure, to be used in magnetoelectric transport studies. The sapphire substrates were cleaned using a modified Radio Corporation of America protocol.[24] The substrates were heated to 820 °C during $V_2O_3$ deposition. A KrF excimer laser with pulse energies of 200 mJ, a spot size of $\approx 6$ mm$^2$, and a pulse width of 20 ns (at a repetition rate of 10 Hz) was used to ablate a $V_2O_3$ target. The target-to-substrate distance was kept at about 9 cm and the substrate rotation rate was at 4 rpm. $Cr_2O_3$ was then grown on the resulting substrates (described above) using PLD, in the presence of a decaborane ($B_{10}H_{14}$) background gas. Deposition takes place in a decaborane ($B_{10}H_{14}$) vapor background pressure of about $1.0 \times 10^{-4}$ Pa. The pulse energy was 190 mJ and the frequency was 10 Hz for $Cr_2O_3$ target. The temperature of the substrate was maintained at 800 °C during deposition of the chromia films. The growth rate of $Cr_2O_3$ ($V_2O_3$) was about 0.025 nm/s (0.017 nm/s), and the thickness was maintained at 200 nm (20 nm). For the neutron depth profile experiments, 300 nm B-doped $Cr_2O_3$ were grown directly on sapphire substrates as these samples did not require bottom electrodes. The structural properties were characterized using X-ray diffraction and the surface roughness was confirmed to be below a root mean square value of 0.25 nm (see Figs. 1S and 2S in supplementary information). We also performed XPS depth profiling. This destructive method required growth of an additional sample which was fabricated to be comparable to the sample used for transport measurements. To this end a 200 nm thick $Cr_2O_3$ film on a 20 nm $V_2O_3$ bottom layer supported on a (0001) sapphire substrate was fabricated. Further details of growth and characterization are provided in our earlier studies.[16]

For electrical transport measurements, a 5 nm thick Pt film was deposited on $Cr_2O_3$ film via DC magnetron sputtering in a vacuum chamber with base pressure of $1 \times 10^{-8}$ Torr ($1.33 \times 10^{-6}$ Pa) and a process pressure of 5 mTorr (0.667 Pa) with an applied power of 30 W. The heterostructure thus consists of a Pt thin film on top of 200 nm B-doped Chromia film grown on a back gate $V_2O_3$ film (20 nm). Pt was then patterned into 1 µm x 8 µm sized star-shaped Hall bar structures using e-beam lithography. Additional details of device fabrication are provided in the supplementary information of Ref.[16]. We note that samples of different thicknesses have been grown specifically for the different measurement techniques. For transport measurements, our routine chromia layer thickness is 200 nm, that serves to apply a relatively lower voltage pulse for switching experiments. For XPS measurements we used a similar sample, i.e., a 200 nm of B:$Cr_2O_3$ on 20 nm $V_2O_3$, supported on sapphire substrate. In the case of cNDP measurements, the instrument resolution required the $Cr_2O_3$ layer to be a few hundred nm and hence 300 nm was chosen to be a nominal value. As the target was to measure B distribution in the sample, it was decided to confine ourselves with using the B:$Cr_2O_3$ directly grown on sapphire to keep the experimental observations straightforward.

## 3. Cold Neutron Depth Profiling

Cold neutron depth profiling experiments were completed at the National Institute of Standards and Technology (NIST) Center for Neutron Research (NCNR).[25] Within this facility cold neutrons are provided by a 20 MW nuclear reactor equipped with a liquid hydrogen cold source.[26] Experimental data were collected at Neutron Guide-5 (NG-5) at the NIST cold neutron depth profiling station (see Ref. [27] for instrument description - note that the instrument location and neutron fluence rate has changed since the publication of this cited manuscript). cNDP experiments were conducted by mounting a sample behind a ≈ 5.0 mm (diameter) opening in a ≈ 5 mm thick Teflon (fluorinated ethylene propylene) circular aperture. This aperture was mounted onto a large Al disk with a ≈ 30 mm (diameter) hole in its center, which was aligned to face a circular transmission-type silicon surface barrier detector (Ametek). The detector was located ≈ 120 mm from the sample surface. NDP results are the average distribution of $^{10}B$ across the surface area of ≈ 5.0 mm aperture opening. $^{10}B$ reacts with cold neutrons in two reaction pathways:

93.57% react according $\quad ^{10}B + n_{cold} \rightarrow\ ^{4}He\ (\alpha, 1472\ keV) +\ ^{7}Li\ (840\ keV)$

(1)

while 6.43% react according $\ ^{10}B + n_{cold} \rightarrow\ ^{4}He^{*}\ (\alpha, 1776\ keV) +\ ^{7}Li^{*}\ (1013\ keV)$

All four charge particle products of the $^{10}B(n,\alpha)^7Li$ reaction were measured, but only the $^4He$ particle profile from the 93.57% branching ratio reaction was utilized to calculate final concentration values.[28] This is due to the high counting statistics of this profile as compared to the other three profiles, as well as the low profile overlap with the low energy background spectrum. Each sample was measured for ≈ 30 h. Profiles of a non-borated chromia film (experimental blank), a $^{10}B$ concentration standard (in-house, N6 series), and a $^{10}B$ energy reference material (NIST SRM93a) were also collected using the same experimental setup. Each area analyzed was irradiated at a near constant cold neutron fluence rate of 1.2× $10^9$ neutrons cm$^2$ sec$^{-1}$, with any variations in the fluence rate being corrected via data collected from a neutron monitor. The data was binned to the approximate resolution of the detector (≈ 20 keV).

The $^{10}B$ atom cm$^{-2}$ concentration was calculated using $[a] = [b]\frac{\sigma_{0b}}{\sigma_{0a}}$ where [a] and [b] are the concentrations ($^{10}B$ atoms/cm$^2$) of isotopes a and b being measured in the sample and standard, respectively, and $\sigma_0$ is the thermal neutron cross-sections for the charged-particle emissions (≈3600 barns, 1 barn = 1x10$^{-28}$ m$^2$). Concentration uncertainty is reported to 1σ and is based on experimental counting statistics.

Conversion of data to depth profiles was completed via construction of a channel vs. energy (keV) and an energy (keV) vs. depth (nm) calibration curves. The data in the channel vs. energy (keV) curve were collected experimentally using a similar setup to that described above. The data in the energy (keV) vs. depth (nm) calibration curve were acquired by modeling α particle transport and interaction with the device's chromia layer in TRIM, a subprogram of SRIM (2008).[29] An assumed theoretical density of 5.22 g cm$^{-3}$ for the chromia ($Cr_2O_3$) layer was used in all modeling along with an initial α particle energy of 1472.42 keV and mass of 4.0015 u (1.6605x10$^{-27}$ kg).[27] Profiles shown in Fig. 1a have the depth scale reported relative to $Cr_2O_3$. The sapphire layer and vacuum were not modeled.

Samples were heat-treated using a top-loading closed-cycle refrigerator (CCR, Janis) and a temperature controller (Lakeshore 340). Only one (1) sample was heat-treated at a time to prevent cross-sample contamination. For heating, a sample was loaded into a clean aluminum holder and then enclosed in the CCR vacuum chamber, which was subsequently evacuated to a pressure of ≈ 0.001 Pa. This pressure was maintained over the course of the heat-treatment. Samples were heated to either 500 K or 400 K depending on the experimental guidelines using a heating rate of 5 K/min. The samples were held at their respective treatment temperatures for ≈ 18 h, after which they were cooled to room temperature using a cooling rate of 5 K/min. Samples were immediately loaded into and measured by NDP following completion of their heat-treatments.

Fig.1a shows the measured B-concentration profile of the as-deposited sample prior to heat treatment. The concentration profile is virtually constant and, as expected, decreases at the interfaces to vacuum (depth = 0 nm) and sapphire (depth ≈ 300 nm) respectively. The concentration depth profile after 18 hours of annealing at 500 K (circle) and a second annealing cycle of 18 hours with total exposure time of 36 hours at 500 K (triangles) are also presented in Fig. 1a. The two data sets from the heated samples are identical within uncertainty indicating that the thermally activated change of the concentration profile reached an equilibrium state after the first annealing cycle. Remarkably, the equilibrium post-annealing concentration profile shows a peaked depth dependence. The B-concentration appears almost all at the surface or just below (up to ≈ 50 nm, which is slightly above the estimated resolution of the detector for the analyzed charged particle) the surface of the film and is virtually zero below a depth of 100 nm. The post-annealing concentration depth profile reveals B-accumulation within a surface layer of ≈ 50 nm thickness resulting in an increase in the B-concentration gradient. Such a phenomenon is inconsistent with a mechanism based on B-diffusion.

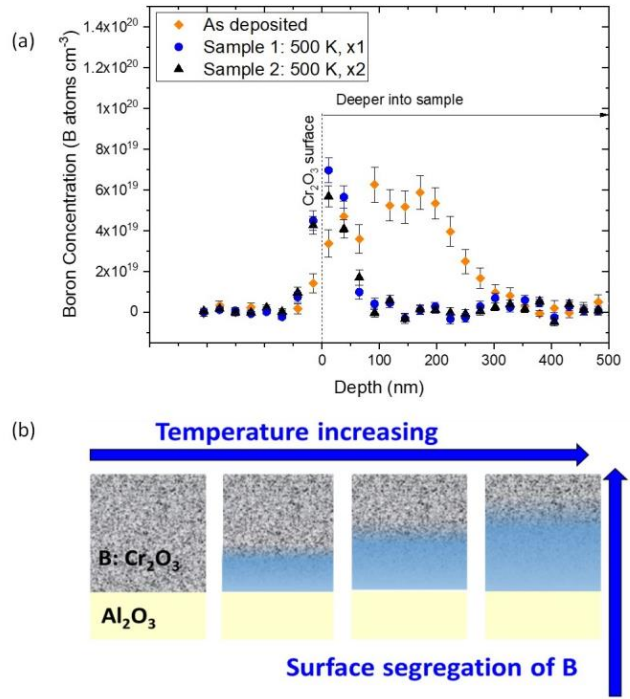

**Fig.1**: (a) B-concentration depth profile of the as-deposited and heated samples measured via cNDP. Depth = 0 nm is the B:$Cr_2O_3$/vacuum interface and depth ≈ 300 nm is the interface between the as-deposited B:$Cr_2O_3$(0001) film and the $Al_2O_3$(0001) substrate (diamonds). Post-annealing B-concentration depth profiles measured at room temperature after 18 hours (circles) and 36 hours (triangles) annealing at 500 K. Data represented to one standard deviation and are based on experimental counting statistics. (b) Schematic depicting surface segregation of B because of annealing. Grey regions are those where B resides in the as-prepared or increased concentration range. Blue regions resemble B-depleted regions.

In the case of diffusion, particle currents are driven by concentration gradients. The evolution of a system towards equilibrium is associated with a reduction of the concentration gradient, which causes the particle current. Fig.1 shows cNDP data sets taken after annealing for 18 hours and 36 hours. These two data sets are virtually identical indicating that the B-concentration increase in the 50 nm surface layer region is a thermodynamic equilibrium feature. A process solely driven by a concentration gradient would have progressed as long as a concentration gradient can drive a particle current. This has not been observed. Rather a stable equilibrium profile evolved. The measured profile, which shows a sizable concentration gradient in thermal equilibrium, is rather the result of a surface segregation process. Surface segregation describes the enrichment of an element (here B) in an alloy at the surface with respect to the bulk. The concentration profile resulting from surface segregation is an equilibrium property and has been extensively studied in metal alloys.[30] It is determined by the temperature dependent Gibbs surface free energies of the constituents. That implies on one hand that the annealing temperature controls the surface segregation and, in contrast to solely concentration drive diffusion, leads to a concentration profile with a gradient near the surface. Although the overall amount of B, integrated over the volume of the sample, might very well be reduced on annealing, it appears from Fig. 1a that the B-concentration in the 50 nm subsurface layer increased as a result of annealing. In this context it is worth to remember that diffusion, as described by Fick's laws, reduces concentration gradients in accordance with Le Chatelier's principle. Surface segregation is a process which, once equilibrium is reached, ensures that the chemical potential across the sample and between bulk and surface of the sample is homogeneous. Although the chemical potential depends on concentration, and a concentration gradient typically contributes to a non-uniform chemical potential, the difference between bulk and surface free energy compensates the concentration dependent contribution to the chemical potential and stabilizes a concentration gradient near the surface. This concentration gradient is necessary to establish homogeneity of the chemical potential and thus thermodynamic equilibrium.

The thermodynamics of surface segregation bears similarities to the thermodynamics of an ideal gas of particles of mass, $m$, at constant temperature, $T$, subject to a gravitational field of constant gravitational acceleration $g$. When defining the particle density, $n$, at the bottom $h = 0$ of the column of gas as $n(h = 0)$, one can introduce the elevation dependent particle density $n(h) = n(h = 0)e^{-mgh/k_BT}$. The $h$-dependent particle density implies the particle density gradient, $\left(0, 0, \frac{dn}{dh}\right) = -\frac{mg}{k_BT} n(h = 0) e^{-\frac{mgh}{k_BT}} \hat{e}_z$. It is the non-zero component of this gradient which ensures that the chemical potential $\mu = k_B T \ln \frac{n(h)}{n_Q} + mgh$ fulfills $\frac{d\mu}{dh} = 0$. This analogy shows that, for a gas in a gravitational field, the equilibrium condition of a constant chemical potential is realized due to the presence of the concentration gradient. The presence of the gravitational field in the gas plays a similar role as the presence of a surface in B:$Cr_2O_3$ films. The surface creates a contribution to the chemical potential which enables the presence of a B-concentration profile in B:$Cr_2O_3$. The B-concentration profile gives rise to a depth invariant chemical potential and thus thermal equilibrium.

The final concentration profile in our sample is stable over time because it is associated with the equilibrium state determined by the minimum of the total Gibbs free energy. The phenomenon of surface segregation (as depicted in Fig. 1b) is not limited to metallic alloys but also is well-known in materials with an electronic band gap such as the III-V semiconductor alloys and transition metal oxides.[31,32] Because it is known that the B-concentration in B:$Cr_2O_3$ determines its Néel temperature,[5,7] it is expected that the B-rich surface layer has a strongly increased Néel temperature compared to $Cr_2O_3$.

## 4. X-ray Photoemission Depth Profiling

Experiments dependent on neutrons generated in a nuclear reactor are expensive and, as such, completed under time limitations. Fig. 1a shows cNDP data for the as prepared sample and depth profiles after annealing at 500 K for 18 hours and 36 hours.

In order to elucidate the role of the annealing temperature in the continuous evolution of the B-concentration profile we independently performed table-top experiments measuring depth profiles with the help of X-ray photoemission spectra (XPS) taken after annealing the sample at increasing temperatures. A sapphire (0001)/$V_2O_3$ (20 nm)/$Cr_2O_3$ (200 nm) sample was grown specifically for the destructive depth profiling measurements.

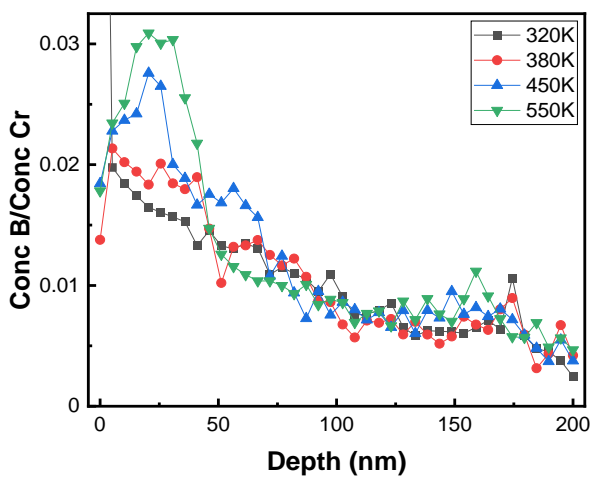

Fig. 2: B-concentration depth profiles of sapphire(0001)/$V_2O_3$ (25 nm)/B:$Cr_2O_3$ (200 nm) measured via XPS. Spectra are taken at room temperature and on subsequent local Ar-ion etching. Depth profiles are measured after annealing the sample at 320 (squares), 380 (circles), 450 (up triangles) and 550 K (down triangles) for 2 hours respectively. The interface between sample surface and vacuum defines depth=0.

Figure 2 shows the results of subsequent XPS measurements using a ThermoFisher K-Alpha Photoelectron Spectrometer. Specifically, we measured the binding energies of the 1s electrons of B and the binding energies of the $2p_{1/2}$ and $2p_{3/2}$ electrons of Cr. Squares, circles, up and down triangles represent XPS data taken at room temperature after annealing the sample at 320, 380, 450 and 550 K. Each data point represents an XPS spectrum taken at a particular depth of the sample with a probing depth of 1nm. For each annealing temperature, the first spectrum is taken at the surface (depth=0). Subsequent spectra are taken after Ar-ion etching a local region of 5 mm x 2mm (representative XPS spectra are provided in the supplementary material) for 30 seconds at ion energies of 3 keV. The relation between etching time and depth is assumed to be linear and gauged via the intensity of the Cr 2p$_{3/2}$ peak which stays virtually constant between surface and substrate and falls off to zero once the V$_2$O$_3$ bottom layer at depth ≈ 200 nm is reached (see supplementary material for depth profile of the Cr 2p$_{3/2}$ peak and details regarding background subtraction). For each data point shown in Fig. 2 we calculated the integrals $I_B(1s)$ and $I_{Cr}(2p_{3/2})$ of the B 1s peak and the Cr 2p$_{3/2}$ peak, respectively, after subtraction of a background modeled with the help of the Shirley function. Calculation of the dimensionless concentration ratio

$$\frac{C(B)}{C(Cr)} = \frac{\Omega_{Cr_{2p}} I_B(1s) E_{kin,B_{1s}}^{0.47}}{\Omega_{B_{1s}} I_{Cr}(2p_{3/2}) E_{kin,Cr_{2p_{3/2}}}^{0.47}}$$

allows for quantitative comparison of the various B-concentration profiles after annealing. Here $\Omega_{Cr_{2p},B_{1s}}$ are the cross-sections for Cr 2p and B 1s and $E_{kin,Cr_{2p_{3/2}},B_{1s}}^{0.47}$ corrects for the fact that the transmission of a hemispherical analyzer changes with the electron kinetic energy of the photo electrons.[33]

Clearly, a very significant gradual increase of B-concentration within a surface layer of about 50 nm width is seen in the XPS spectra in agreement with the neutron data and in agreement with the interpretation of the transport data. In addition to the evolution of the pronounced peak near the surface, the XPS depth profiles also reveal a gradual increase in B-concentration from the interface with the substrate to the surface. We interpret this gradual increase as B-migration taking place during the growth process of the sample. A similar observation holds for the cNDP data of the as prepared sample where B-concentration gradually increases from the sapphire interface at d ≈ 300 nm to its maximum value at d≈ 200 nm. In contrast to Fig. 1a, which shows only the difference in the B-concentration between the as prepared sample and the B –concentration profile after annealing at 500 K, the XPS depth profiling displays the gradual effect of annealing on the concentration profile. Note that the sample prepared for XPS measurements has a 20 nm V$_2$O$_3$ bottom layer resembling the structure of the sample used for better comparison with the transport measurements. The absence of an increased B-concentration at the V$_2$O$_3$/B:Cr$_2$O$_3$ interface originates from the structural and dielectric similarity between V$_2$O$_3$ and Cr$_2$O$_3$. The same holds for Al$_2$O$_3$ and Cr$_2$O$_3$. The similarity in structure and dielectric properties between these three sesquioxides and the epitaxial growth of the films minimizes the interface contribution to the chemical potential and hence eliminates the stabilizing effect of the concentration gradient observed near *d*=0 at the vacuum or metallic interface.

Despite the qualitative agreement between the cNDP and XPS depth profiling measurements regarding the formation of a B-concentration increase in a 50 nm layer near the surface, there are striking differences which ask for an interpretation. The cNDP data in Fig. 1 show a clear difference in the Boron concentration in the 100-200 nm range for the as deposited and annealed sample. In the XPS depth profiling data of Fig. 2 this difference is far less pronounced. Both XPS and neutron capture techniques have different probing depths and thus can provide similar trends but cannot provide identical depth profiles. Neutron depth profiling is based off $^{10}$B capture and can be considered as more sensitive than XPS because of low cross-section of B1s core in XPS. Moreover, during XPS measurements, a small rectangular area of the sample was sputtered each time, prior to taking the XPS data. This is a key difference compared to the cNDP. The sputtering creates defects and intermixing as it removes material. The information obtained by XPS at each depth is coming from a 1 nm thick layer where the aforementioned layer mixing can make the experimental error more pronounced at greater depths. Furthermore, the sputtering removes different species at different rates because boron has a higher sputtering cross-section than chromia.[34] In addition, it makes a difference to have a thermodynamically stable surface segregation layer as in the case of the cNDP measurements or a freshly created surface where segregation did not happen and boron can possibly escape.

To take into account the differences in resolution of the two methods, we employed a convolution technique followed by the work by J.F. Ziegler et.al.[35] We display and discuss the result in the supplementary material (Figs. 9S and 10S).

5. Magnetotransport measurements

Hall measurements were taken by sending a D.C. current of 4 µA through one of the legs of a Hall cross and measuring the transverse voltage $V_{xy}$ across the perpendicular leg. Pulses of positive or negative gate voltage, $V_G = \pm 55$ V with duration of $\Delta t = 0.5$ s were applied across the AFM film using the $V_2O_3$ film as bottom electrode and the Pt Hall bar as top electrode (see Fig. 3d for a schematic of the device). Our DC measurements use the Delta mode of the Keithley nanovoltmeter. A set of 100 measurements takes about 2 minutes. The integration time of the measurements electronics is about 17 ms.

Fig. 3a shows $V_{xy}$ for 800 sequential Hall measurements taken at $T = 295$ K with voltage pulses of positive or negative polarity applied every 100 data points. The applied voltage creates an $E$-field ($\approx \pm 55\ V / 200\ nm = \pm 275\ MV/m$) across the B:Cr$_2$O$_3$-film of the device which gives rise to electrically controlled spin manipulation in the AFM film.

Fig. 3b visualizes the pulse train applied during the 800 Hall measurements. Vertical dashed lines correlate $E$-pulses with the $V_{xy}$-signal shown in Fig. 3a. Unlike our previous work,[16] here we go beyond a scheme of pulses with periodically alternating polarity. The more intricate protocol applied here allows to provide evidence for the polarity dependence and deterministic nature of the Néel vector rotation. Initially, a pulse of $V_G = -55$ V is applied. The subsequent 100 Hall measurements show that a state with $V_{xy} \approx 0$ has been initialized where Néel vector and boundary magnetization are oriented in-plane. Prior to the 101$^{st}$ Hall measurement, a pulse of $V_G = +55$ V is applied. It rotates Néel vector and boundary magnetization out of plane giving rise to $V_{xy} \approx 7$ mV. Next, $V_G = -55$ V is applied prior to the 201st data point. This pulse brings the AFM back to the $V_{xy} \approx 0$ state. Evidence for the polarity dependence of the rotation is provided by applying again a pulse of $V_G = -55$ V prior to the 301st Hall measurement. Because the system is already in the $V_{xy} \approx 0$ state, the pulse of identical polarity to the previous one leaves the AFM in the $V_{xy} \approx 0$ state.

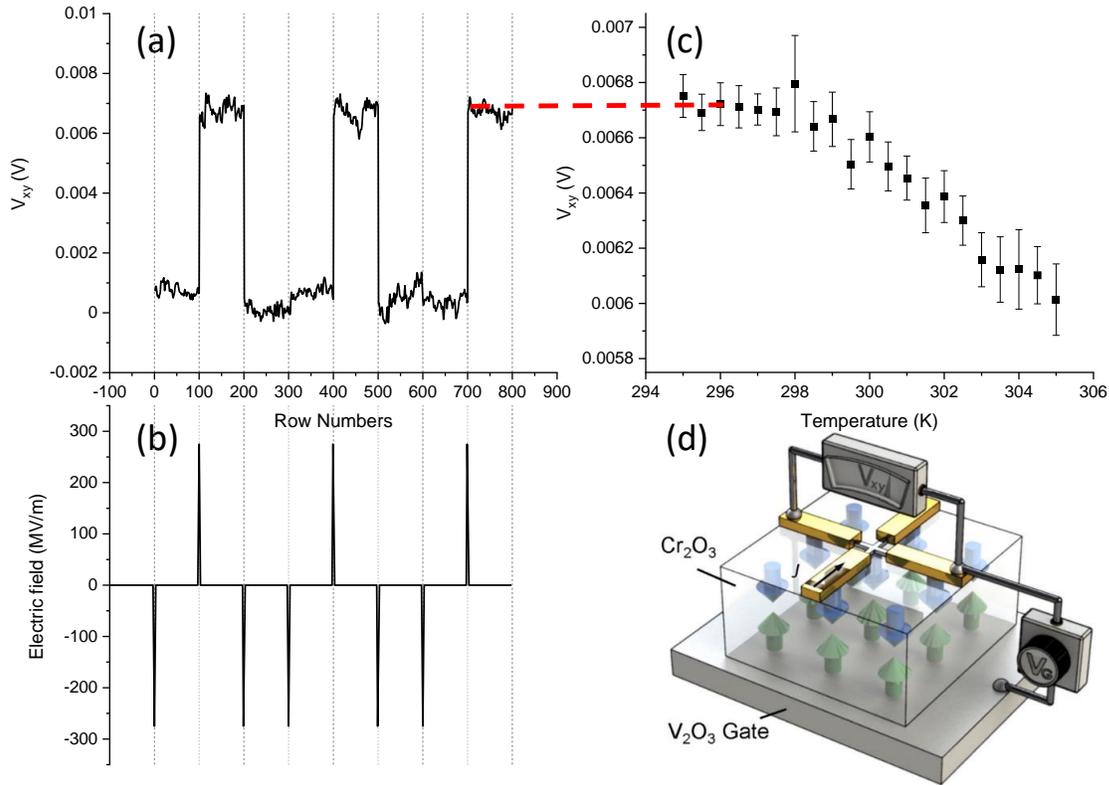

**Fig. 3**: Electric field driven manipulation of the AFM order parameter and the Néel vector orientation. (a) $V_{xy}$ switching between a higher and zero voltage at 295 K, corresponding to the application of gate pulse voltage as shown in (b). (c) $V_{xy}$ measured as a function of increasing temperature after the application of $V_G = +55$ V at 700th measurement. Dashed horizontal line indicates that the $T$-dependent measurement starts at the $V_{xy}$-value set by the last gate pulse. Uncertainty bars represent the standard deviation calculated for each set of 100 $V_{xy}$ measurements at each temperature. (d) Cartoon of Hall bar device showing $V_2O_3$ back gate, B-doped $Cr_2O_3$ film with AFM spin structure, Pt Hall cross with Au electrodes, current density $j$ flowing in direction of black arrow causing signal $V_{xy}$ controlled by gate voltage $V_G$ (adapted from Mahmood et al., 2021[ref 16], not to scale).

Similarly, $V_G$ = +55 (-55) V applied prior to the 400th (500th) point brings the AFM in a state of non-zero (zero) $V_{xy}$ while a pulse of unchanged negative polarity prior to the 600th point leaves the AFM state unchanged at $V_{xy} \approx 0$. Thus, a completely deterministic polarity dependent voltage switching is demonstrated. A $V_G$ = +55V pulse applied prior to the 700th measurement selects the $V_{xy} \approx 7mV$ state associated with out of plane orientation of boundary magnetization and Néel vector. Immediately after application of this pulse and without exposure to annealing, the temperature is increased from $T_1$= 295 K to $T_1^{max}$=305 K and Hall measurements are taken.

Fig. 3c shows $V_{xy}$ *versus* $T$ measured on increasing the temperature from $T_1$ to $T_1^{max}$. $V_{xy}(T_1)$ coincides with the value of $V_{xy}$ set by the last gate pulse and decreases with increasing temperature. The horizontal dashed line between the ordinates of Fig. 3a and Fig. 3c highlights the fact that the $T$-dependent measurement starts at the $V_{xy}$ value associated with out of plane orientation of the Néel vector. The decrease of $V_{xy}$ *vs*. $T$ is in accordance with the expected $T$-dependence of the AFM order parameter and the associated boundary magnetization, which both vanish at $T_N$. Qualitatively, this $T$- dependence is well-known from $Cr_2O_3$-based Hall-bar structures where it is straightforward to measure $V_{xy}$ *vs*. $T$ at temperatures near and above $T_N$. [12,21,36] The situation changes in Hall-bar structures based on B: $Cr_2O_3$. Here annealing alters the B-concentration profile whenever the sample is exposed to a new maximum temperature for the first time. The increase in B-concentration at the B: $Cr_2O_3$/Pt interface increases the surface layer Néel temperature. To keep the change in the B-concentration in a given measurement cycle small and to systematically map the evolution of $T_N$ on annealing, the maximum temperature, $T_n^{max}$, of the $n$th measurement within [295 K, $T_n^{max}$] is kept significantly below the surface Néel temperature, $T_N(T_{n-1}^{max})$, induced by the previous measurement within [295 K, $T_{n-1}^{max}$] where $T_{n-1}^{max} < T_n^{max}$.

$T_N(T_n^{max})$ is determined from a scaling analysis outlined in detail below. Within a single temperature sweep, the data can be modeled in good approximation with the help of one constant value of $T_N$. Because of the $T$-induced change in the B-concentration profile, the virgin $V_{xy}$ *vs*. $T$

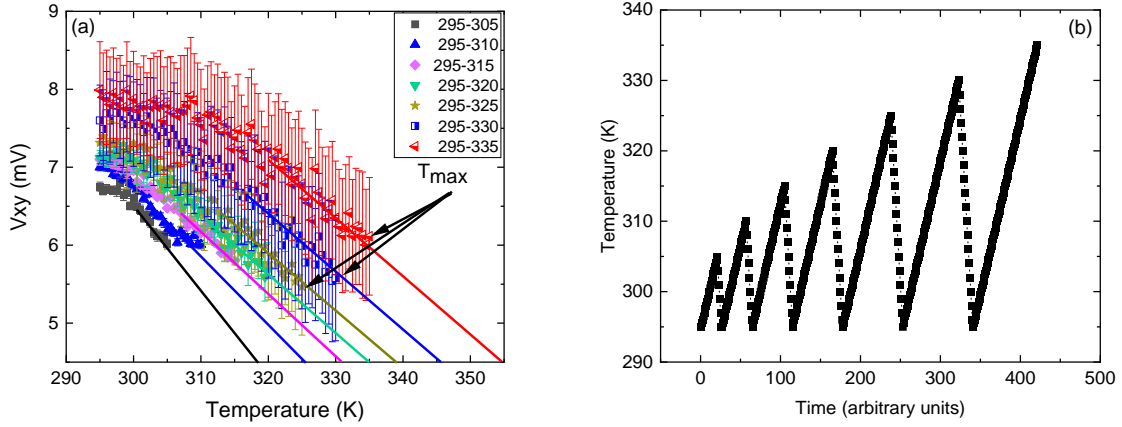

**Fig 4**: (a) $T_N$-runaway effect shown in the evolution of $V_{xy}$ vs. $T$ for temperature intervals $T_1 = 295K \leq T \leq T_n^{max}$ with $T_1^{max} = 305\,K$ (squares), $T_2^{max} = 310\,K$ (solid up triangles), $T_3^{max} = 315\,K$ (solid squares), $T_4^{max} = 320\,K$ (solid slanted triangles), $T_5^{max} = 325\,K$ (stars), $T_6^{max} = 330\,K$ (semi filled squares), and $T_7^{max} = 335\,K$ (semi-filled triangles). Uncertainty bars represent the standard deviation calculated for each set of measurements at each temperature. (b) Schematics of the temperature ramps for measurements shown in Fig. 4a. The temperature is linearly increased in steps of 0.5 K and at each step stabilized for 10 minutes.

measurement cannot be reproduced. After completion of a measurement, the $T_N$ of the surface layer has changed to a new equilibrium value such that $T_N(T_n^{max}) > T_N(T_{n-1}^{max})$.

To systematically investigate this $T_N$-runaway effect, the following experimental protocol is applied. $V_{xy}$ vs. $T$ is measured for temperature intervals $T_1 = 295K \leq T \leq T_n^{max}$ of increasing width with $n= 1, 2, 3, \ldots$ and $T_{n+1}^{max} = T_n^{max} + 5K$ while $T_1^{max} = 305\,K$. That means, the $n^{th}$ data set is comprised of $V_{xy}$ vs. $T$ data taken on heating from $T_1 = 295\,K$ to $T_n^{max} = 305\,K + (n - 1) \times 5K$.

Data sets associated with this protocol are shown in Fig. 4 (additional data at higher temperature $T_n^{max}$ are shown in Fig. 11S in the supplementary material). The eye-guiding straight lines qualitatively show that $T_N$, defined by $V_{xy}(T \geq T_N) = 0$, increases systematically with increasing annealing temperature $T_n^{max}$. In addition to the increase of $T_N$, the data show that $V_{xy}(T, T_{n+1}^{max}) > V_{xy}(T, T_n^{max})$. This behavior is consistent with the fact that the magnitude of the AFM order parameter $\eta \propto V_{xy}$ is determined by the distance $\Delta T = T - T_N$ as expressed in the functional form $\eta \propto (T - T_N)^\beta$ valid near $T_N$. Evidently, the higher the $T_N$ the larger is $\eta$ at a given temperature $T < T_N$. Qualitatively, this holds not only near the critical temperature but for all $0 < T \leq T_N$.

## 6. Scaling analysis and evolution of the surface $T_N$ on annealing

The $T$-dependent surface segregation of boron and the associated $T_N$ -runaway effect, *i.e.*, $T_N(T_n^{max}) > T_N(T_{n-1}^{max})$, make it challenging to measure $V_{xy}$ vs. $T$ near $T_N$. To determine $T_N$ in the absence of data near $T_N$, we employ a scaling approach. It allows to estimate $T_N(T_n^{max})$ from $V_{xy}$ vs. $T$ data taken within $T_1 = 295K \leq T \leq T_n^{max}$ where $T_n^{max}$ is significantly below $T_N(T_n^{max})$.

The scaling analysis utilizes the solution of the mean-field equation which approximates the $T$-dependence of spontaneous order below a critical temperature. In contrast to the power-law $\eta \propto (T - T_N)^\beta$, which describes critical behavior with the help of the universality class dependent critical exponent, $\beta$, the mean-field approximation is independent of dimensionality and symmetry of a spin-Hamiltonian. It does not discriminate between universality classes but has the advantage of being a meaningful approximation outside the critical region, *i.e.*, away from the critical point. Microscopic details, such as the number of nearest neighbor spins and their interaction strength, can be absorbed in the critical temperature as a single parameter. In the absence of an applied conjugate field the mean-field equation reads[37]

$$\eta(T) = \eta_0 \tanh\left(\frac{T_N}{T} \eta\right), \qquad (2)$$

where $\eta_0 = 1$ in case $\eta$ describes a normalized order parameter. $\eta_0 \neq 1$ considers a saturation value when $\eta$ refers to a quantity proportional to the order parameter such as $V_{xy}$.

The solid line in Fig.5a shows the numerical solution of Eq.(2). When plotted in the format $\eta/\eta_0$ *vs.* $T/T_N$, each data set that can be approximated by Eq.(2), will fall onto this universal curve (line in Fig.5a) regardless of the specific value of $T_N$. The position of the data on the universal curve uniquely determines the critical temperature $T_N$ associated with a data set. We utilize the scaling approach to estimate $T_N(T_n^{max})$ from the collapse of the $V_{xy}$ *vs.* $T$ data on the mean-field master curve. For quantitative optimization of the data collapse we set up the functional

$$S(V_{0,n}, t_n) = \sum_{i=1}^{K_n} \left(\frac{V_{xy}(T_i)}{V_{0,n}} - \tanh\left(V_{xy}(T_i) * \frac{t_n}{T_i}\right)\right)^2. \qquad (3)$$

It assigns the positive number $S(V_{0,n}, t_n)$ to the $n^{th}$ data set, which contains $K_n$ data points $V_{xy}(T_i)$ within $T_1 \leq T_i \leq T_{K_n} = T_n^{max}$. $S(V_{0,n}, t_n)$ quantifies the deviation of $V_{xy}(T_i)$ from perfect collapse ($S=0$) on the master curve. Minimization of $S(V_{0,n}, t_n)$ with respect to $V_{0,n}$ and $t_n$ determines the critical temperature $T_N(T_n^{max}) = t_n V_{0,n}$. When plotting $\frac{V_{xy}^i}{V_{0,n}}$ vs. $\frac{T}{t_n V_{0,n}}$, the best possible data collapse based on the least square criterion of Eq.(3) is achieved.

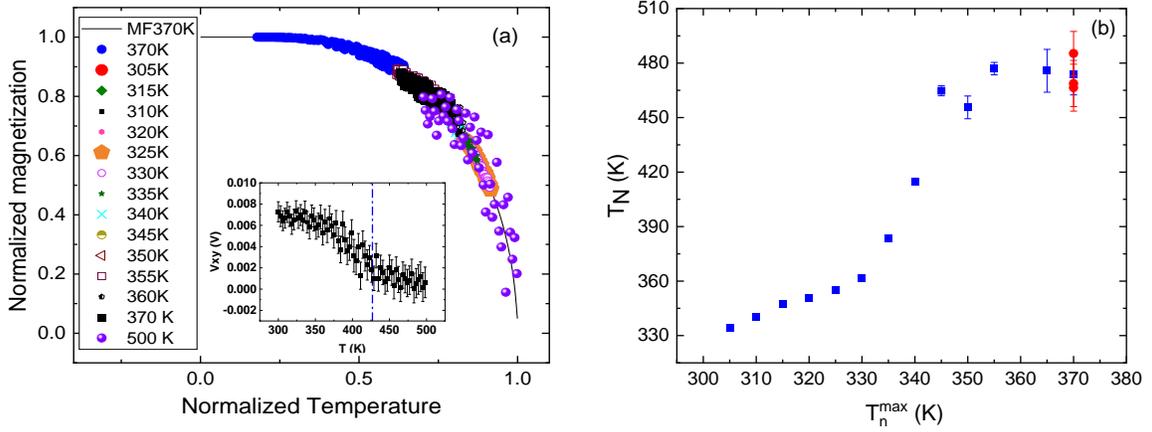

**Fig 5**: (a) Scaling analysis of $T$-dependent magnetotransport data. Uncertainty of these data sets are displayed in Fig. 4 and the inset. Line shows the numerical solution of the mean-field Eq.(2) which serves as master curve for a data collapse of $V_{xy}$ vs. $T$ data. The legend displays symbols associated with $V_{xy}$ vs. $T$ and their $T_n^{max}$ (see text for details). The inset shows a complete $V_{xy}$ vs. $T$ data set which includes data for $T>T_N$. Data were taken three months after the sample has been exposed to $T_{max}=335$ K. Vertical line marks $T_N$ as determined from scaling analysis. (b) Plot of $T_N$ vs. $T_n^{max}$ (squares) where the respective $T_N$-value has been determined by minimizing the functional of Eq.(3). Circles show data experiments taken with a time delay of weeks after the annealing procedure. Uncertainty bars are determined by repeating the analysis for data sets with variable sampling size of data points. The uncertainty bar of a respective $T_N$-value indicates the standard deviation determined from 5 data sets of different sampling sizes.

Fig.5 a shows the virtually perfect data collapse of all $V_{xy}$ vs. $T$ data sets. The legend in Fig.5a provides the values of $T_{K_n} = T_n^{max}$ while $T_1 = 295\,K$. This is the case for all data sets with the exception of the solid blue circles extending down to $T/T_N = 0.2$. Here data taken between 100 K and 295 $K$ are combined with data taken between 295 $K$ and 370 K and $S(V_{0,n}, t_n)$ is minimized for the combined data. Note that the data between 100 K and $T_1$ have been measured after the sample has been exposed to 370 K. Once the sample has been annealed at temperatures above 345 K, $T_N$ vs. $T_n^{max}$ saturates (see Fig. 5b). The saturation value $T_N^{sat} \approx 477\,K$ becomes independent of subsequent temperature protocols.

Fig. 5b shows $T_N$ vs. $T_n^{max}$ as obtained from the scaling analysis. Uncertainty bars are determined by repeating the analysis for data sets with variable sampling size of data points. The uncertainty bar of a respective $T_N$-value indicates the standard deviation determined from 5 data sets of different sampling sizes. For the as-deposited sample we find $T_N \approx 334\,K$ (uncertainty bar smaller than the plot symbol). On annealing, that is with increasing $T_n^{max}$, $T_N(T_n^{max})$ increases linearly up to $T_N(T_{n=6}^{max} = 330\,K) \approx 362\,K$. Only when $T_n^{max}$ reaches the Néel temperature of the as prepared sample ($T_N \approx 334\,K$), the annealing effect becomes sizable with a pronounced increase from $T_N(T_{n=7}^{max} = 335\,K) \approx 383\,K$ to $T_N(T_{n=9}^{max} = 345\,K) \approx 465\,K$. For $T_{n>9}^{max} > 345\,K$ the surface Néel temperature remains virtually constant at $T_N^{sat} \approx 477\,K$. The saturation $T_N(T_{n>9}^{max}) \to T_N^{sat} \approx 477\,K$ and the subsequent history independence of $T_N^{sat}$ corroborate the interpretation of the depth profiling in terms of surface segregation and exclude a B-diffusion mechanism. The latter would continue as time progresses and intensify with increasing annealing temperature until the B-concentration is depleted. Instead, stabilization of a high surface Néel temperature is observed with beneficial implication on device applications.

The main goal of this work is to demonstrate qualitative effects. The transport data, the neutron data, and the XPS data had to be taken at different samples due to the fact that all experiments alter the sample irreversibly. One cannot expect perfect quantitative agreement when comparing the results of three different samples. As can be seen from the analysis of the transport data in Fig. 5b, the annealing effects set in gradually but speed up dramatically in a rather narrow temperature window. It can be expected that this temperature window varies from sample to sample due to factors including variation in doping concentration and defect concentration. Within this context we find the agreement between the three very different methods satisfactory but acknowledge that they remain on a qualitative level.

## 7. Robust post-annealing switching behavior

The stabilization of the post-annealing surface Néel temperature at $T_N^{sat} \approx 477\,K$ suggests that annealing improves and stabilizes device properties. To confirm this expectation, we perform post annealing isothermal voltage-controlled rotation of the Néel vector. Rather than rotating the Néel vector near room temperature as shown in Fig. 3, we perform the Hall measurements at 370 K. In addition, to provide evidence that the device is not degrading over time, we perform the post-annealing switching experiments with a time delay of weeks after the annealing procedure.

Fig. 6a shows deterministic switching at $T = 370$ K, as it follows the voltage pulse application in the same order as the 295 K experiment. The high voltage state of the transverse Hall signal of $V_{xy} \approx 4\ mV$ is measured after application of a gate pulse of $V_G = +45\ V$ (+225MV/m). The zero state is set after application of $V_G = -45\ V$ (-225 MV/m). Note that a lower switching voltage of $\pm 45$V achieves switching at 370 K compared to $V_G = \pm 55$V required for switching at 295 K.

In pure chromia, magnetoelectric switching can be brought about at lower temperatures by applying higher voltage pulses at a fixed magnetic field.[13,14] There are competing mechanisms causing the temperature dependence of the switching field product in pure $Cr_2O_3$. On one hand, the magnetic anisotropy, which determines the energy barrier for switching, increases with decreasing temperature. On the other hand, the parallel magnetoelectric susceptibility, $\alpha_\parallel$, of $Cr_2O_3$ has a pronounced temperature dependence. It peaks around 280 K and becomes small for temperatures far below the Néel temperature and very close to the Néel temperature. As a result, the magnetoelectric change in the free energy, $\Delta F = 2\alpha_\parallel EH$, which must overcome the magnetic anisotropy energy on switching, becomes strongly temperature dependent and particularly small at low temperatures and very close to the Néel temperature. In B: $Cr_2O_3$ the mechanism of Néel vector rotation is different. Rotation takes place in zero applied magnetic $H$-field and the temperature dependence of $\alpha_\parallel$ becomes irrelevant. The reduced switching voltage solely reflects the reduced magnetic anisotropy associated with increased temperature. Reduced switching voltages are highly beneficial for applications. Further reduction of the switching voltage can be achieved by reducing the thickness of the B: $Cr_2O_3$ because the electric field scales inversely proportional to the film thickness. In pure $Cr_2O_3$, the authors achieved switching of the Hall signal via magnetoelectric annealing in films as thin as 20 nm.[11] Extrapolating to B: Cr$_2$O$_3$ – based devices suggests voltage-controlled switching in *H*=0 for gate voltages as low as 4.5 V and possibly lower due to the superior dielectric properties of B-doped chromia, potentially allowing the growth of leakage free samples below 20 nm thickness.

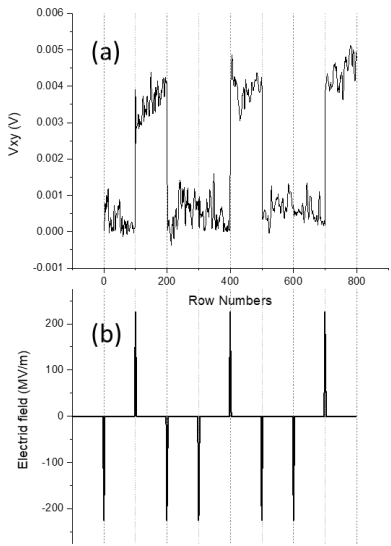

**Fig. 6**: (a) Isothermal voltage-controlled switching between high and zero Hall voltage signals at 370 K, corresponding to the application of gate pulse voltage as shown in (b).

## 8. Conclusions

B-doping of the magnetoelectric antiferromagnet Cr$_2$O$_3$ increases its magnetic ordering temperature and enables pure voltage control of the Néel vector. In this work, we have shown with the help of cold neutron and x-ray photoemission spectroscopy depth profiling that the B-concentration profile in B: Cr$_2$O$_3$ films changes on annealing. Surface segregation gives rise to B-accumulation in a layer of about 50 nm depth where the Néel temperature is increased accordingly. Magnetotransport measurements allow to detect the voltage-controlled rotation of the Néel vector. Scaling analysis of the temperature dependent Hall signal provides the dependence of the surface Néel temperature on the annealing temperature. The surface Néel temperature increases from the as prepared value of $T_N \approx 334\ K$ to a saturation value of $T_N^{sat} \approx 477\ K$ for annealing temperature above 345 K. Isothermal switching experiments performed post annealing demonstrate Néel vector rotation at 370 K, significantly above room temperature. The beneficial device properties, which include high operation temperature with potential application for non-volatile memory and logic in CMOS environments as well as reduced switching voltage, are stable over time due to the long-term stability of the thermally imprinted B-concentration profile.

**Acknowledgement**

This work was supported by the National Science Foundation through EPSCoR RII Track-1: Emergent Quantum Materials and Technologies (EQUATE), Award OIA-2044049. The research was performed in part in the Nebraska Nanoscale Facility: National Nanotechnology Coordinated Infrastructure and the Nebraska Center for Materials and Nanoscience, which are supported by NSF under Award ECCS: 2025298, and the Nebraska Research Initiative. The authors like to thank E. Mishra for her guidance in analyzing of XPS spectra and N. Hong from the J.A. Wollam company for analyzing our samples via ellipsometry. Certain commercial equipment, instruments, or materials (or suppliers, or software, ...) are identified in this paper to foster understanding. Such identification does not imply recommendation or endorsement by the National Institute of Standards and Technology, nor does it imply that the materials or equipment identified are necessarily the best available for the purpose. We acknowledge the support of the National Institute of Standards and Technology, U.S. Department of Commerce, in providing the neutron research facilities used in this work. We would like to thank Tanya Dax and Alan Ye from NIST for helping with the CCR heating treatments.

# Supplementary information

# Post deposition interfacial Néel temperature tuning in magnetoelectric B:Cr$_2$O$_3$

Ather Mahmood[1], Jamie Weaver[2], Will Echtenkamp[1,3], Syed Qamar Abbas Shah[1], Jeffrey W. Lynn[4], Christian Binek[1,*]

[1]*Department of Physics & Astronomy and the Nebraska Center for Materials and Nanoscience, University of Nebraska-Lincoln, Lincoln, NE 68588-0299, USA*

[2]*Material Measurement Laboratory, National Institute of Standards and Technology, Gaithersburg, Maryland, USA 20899*

[3]*Department of Electrical & Computer Engineering, University of Minnesota, Minneapolis, MN 55455, USA*

[4]*NIST Center for Neutron Research, National Institute of Standards and Technology, Gaithersburg, MD 20899, USA*

*To whom correspondence should be addressed. E-mail: cbinek@unl.edu

Supplementary note 1

Fig 1S shows the XRD $\theta$-$2\theta$ patterns for the B-doped chromia sample used for transport measurements in the manuscript. The sample comprises a 200 nm B:Cr$_2$O$_3$ on the 20 nm bottom V$_2$O$_3$ electrode, supported on sapphire substrate. The peaks corresponding to each material are shown in the figure and appear distinctly at 38.5° (V$_2$O$_3$), 39.7° (B:Cr$_2$O$_3$) and 41.7° (Al$_2$O$_3$).

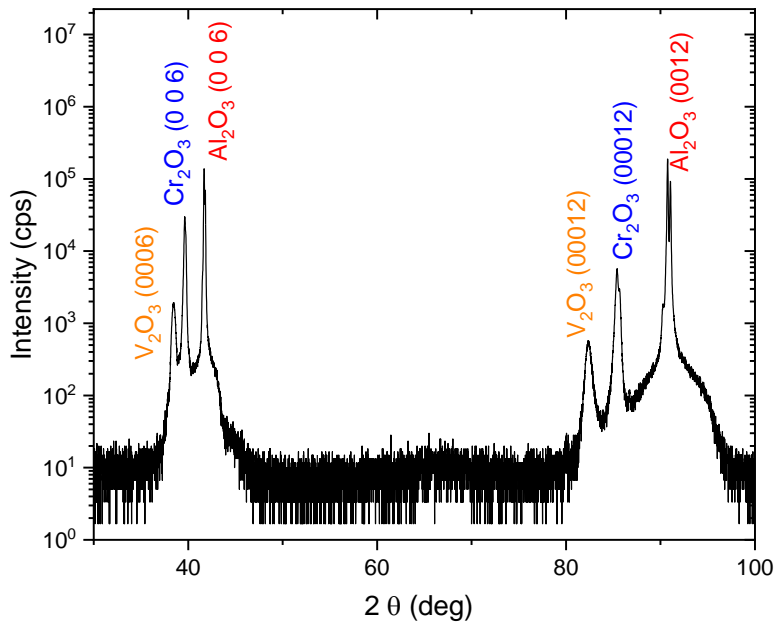

Supplementary note 2

Fig 2S shows a 0.7 µm x 2 µm Atomic force microscope (AFM) image showing the topography of the PLD grown top B:$Cr_2O_3$ layer of the heterostructure discussed above. The box of 500 x 500 nm size represents the area used to analyze the root mean square (RMS) surface roughness of 0.22 nm.

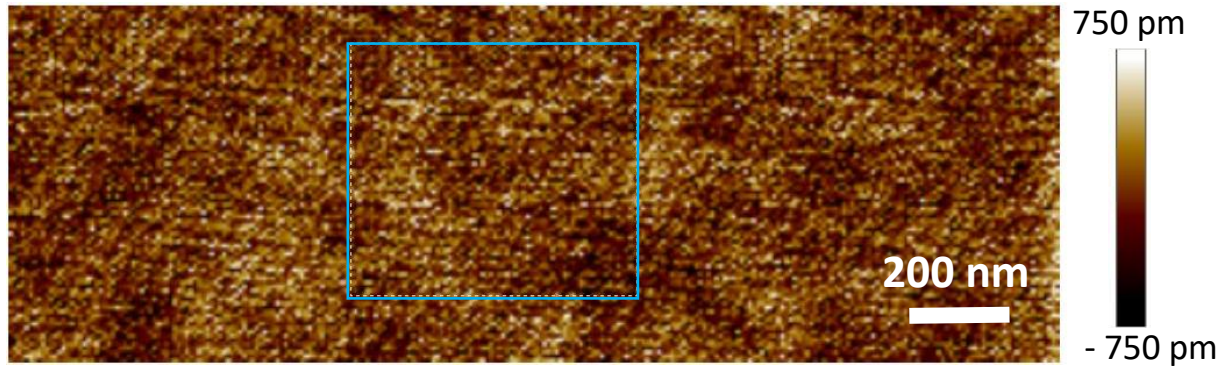

Supplementary note 3

Fig 3S shows the XRD $\theta$-$2\theta$ patterns for the B-doped chromia sample used for neutron scattering experiments, and was captured using a X-ray facility with Co target. The sample comprises a 300 nm B:$Cr_2O_3$ supported on sapphire substrate. The peaks corresponding to each material are shown in the figure and appear distinctly at 45.95° (B:$Cr_2O_3$), and 48.85° ($Al_2O_3$).

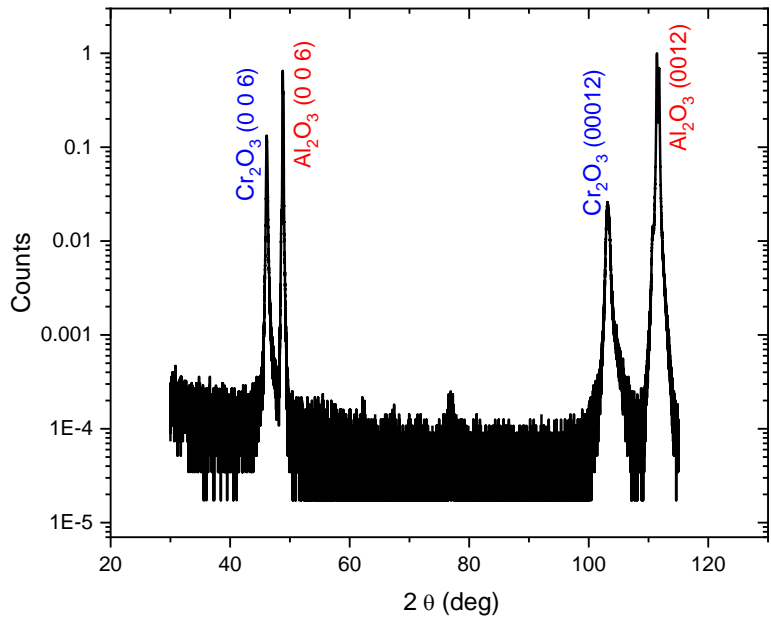

**Dielectric properties**

The leakage current density through B:$Cr_2O_3$ while applying the 55-V gate bias is about $4.3 \times 10^{-5}$ A/$\mu m^2$. The resistivity of 20 nm thin $V_2O_3$ film on sapphire is ~ $10^{-3}$ $\Omega$-cm.

**Supplementary data on XPS depth profiling**

The B doped Chromia thin film sample with a total thickness of 200 nm on top of 20 nm Vanadium Oxide bottom electrode was etched via Ar-ion etching. The X-ray photo emission spectroscopy (XPS) spectra was first taken at the surface of sample d=0 nm. For the XPS depth profile the film was then exposed to the Ar-ions with energies of 3000 eV for 30 seconds at a local region of 5 mm (x-direction) x 2 mm (y-direction). The XPS depth profile spectra were taken after around every 30 seconds. **Figure 1 (a-d)** shows the Chromia 2p peaks of the sample after annealing it at a temperature of 320 K for 2 hours. The complete 200 nm film was etched by bombarding Ar-ions for a total time of 1170 seconds giving the etching rate of 0.17 nm/sec for the B: Chromia.

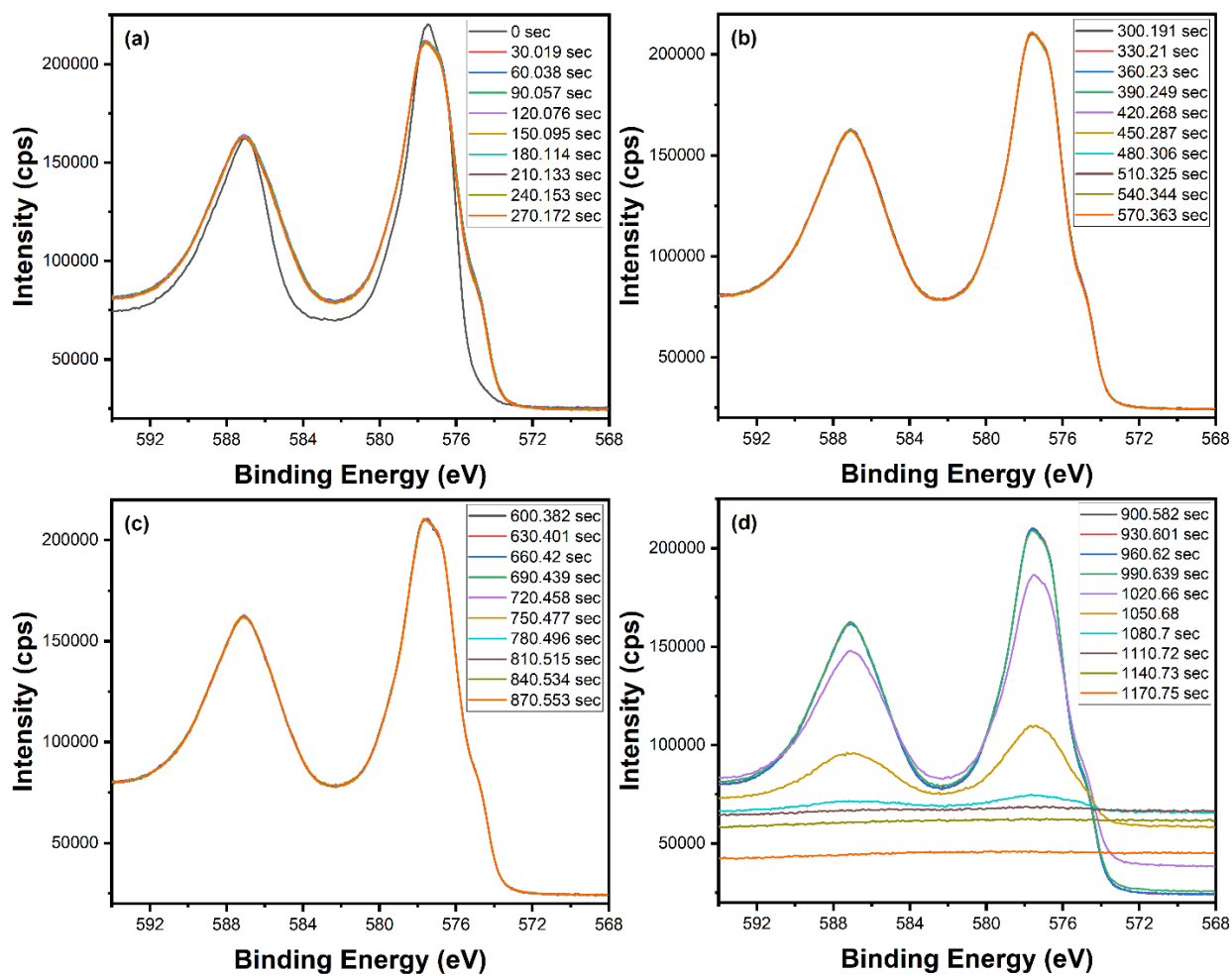

**Figure 4S (a-d)** *Shows the Chromia 2p peaks of the B: Chromia sample after annealing it at a temperature of 320K for 2 hours. The Cr $2p_{1/2}$ appears in the binding energy range of 587 eV and Cr $2p_{3/2}$ appears in the binding energy range of 578 eV.*

**Figure 5S (a-d)** shows the XPS depth profile data for B 1s peak after annealing the sample at a temperature of 320 K for 2 hours.

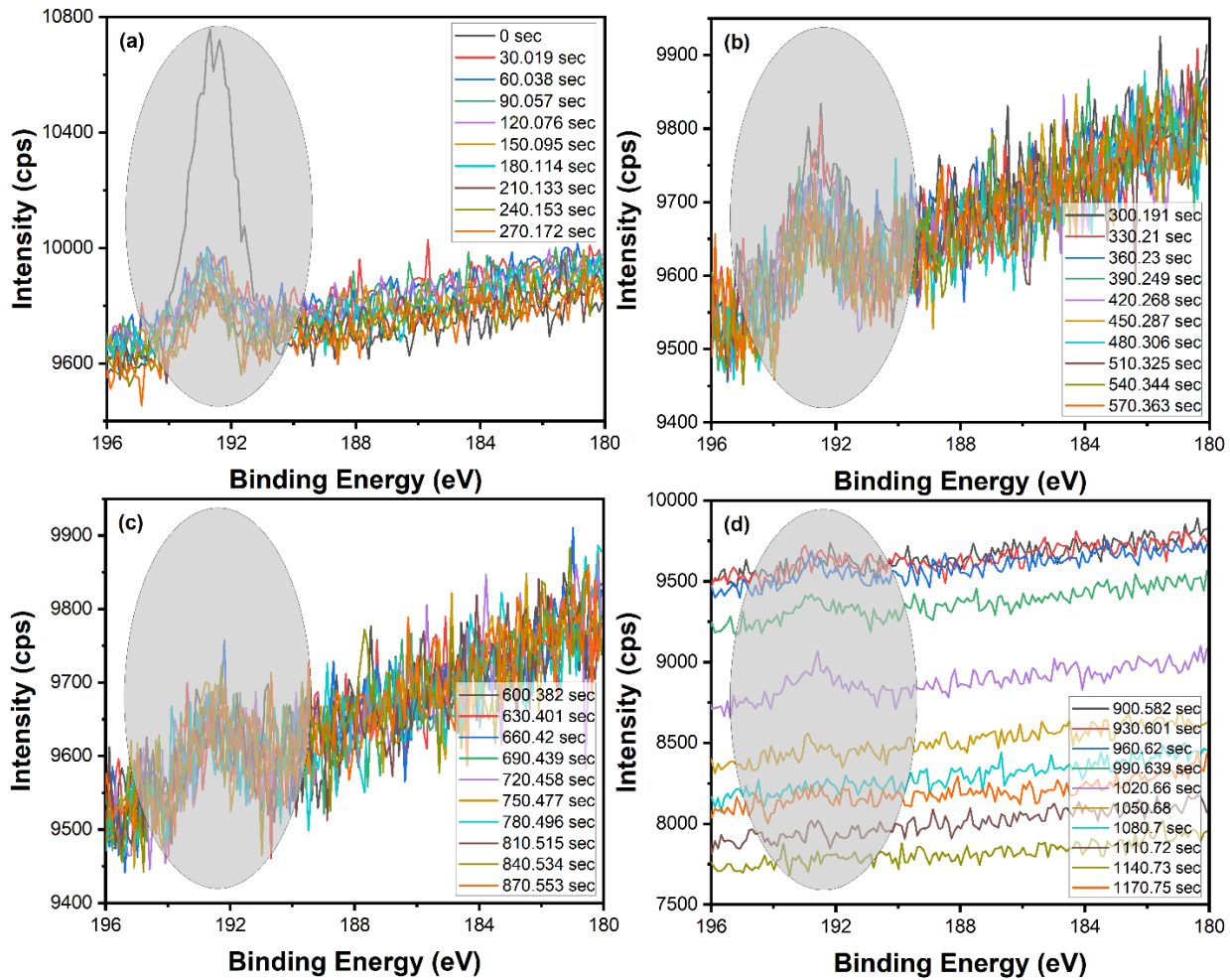

**Figure 6S (a-d)** *Shows the Boron 1s peaks of the B:Chromia sample after annealing it at a temperature of 320 K for 2 hours. The shaded grey region shows the B 1s peaks that appear in the binding energy range of 192.6 eV.*

The XPS depth profile spectra were taken by annealing the sample in heating furnace at temperatures of 320 K, 380 K, 450 K and 550 K for every 2 hours. The data was analyzed by using CasaXPS software. The intensity for every induvial peak was calculated after fitting the data by subtracting the background using the Shirley approach. **Figure 7S-a (a-f) & Figure 7S-b (g-l)** shows the fitted XPS spectra for B 1s peaks after background subtraction for the sample annealed at 550 K.

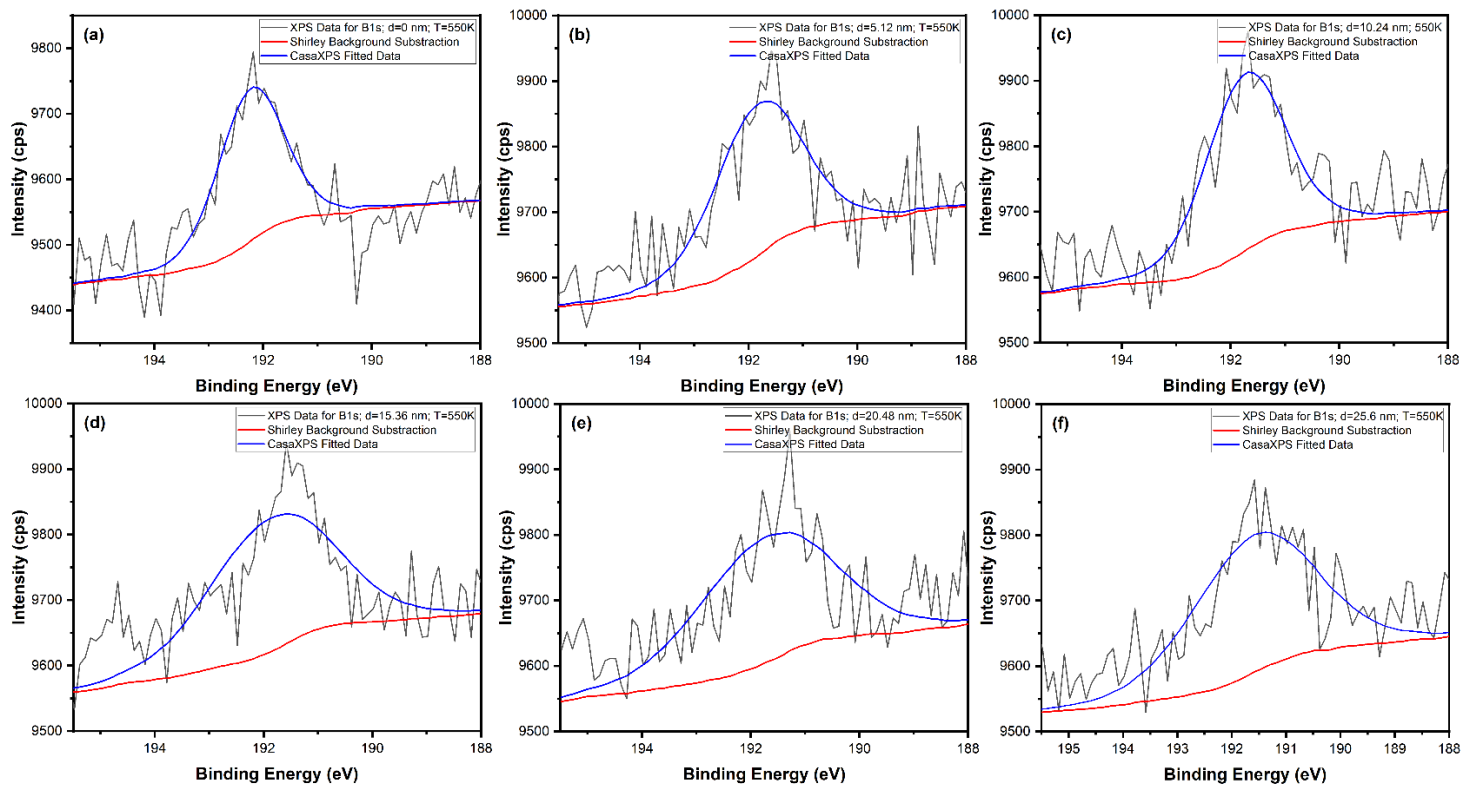

*Figure 3-a (a-f): Figure shows the XPS data for B 1s peaks (black solid lines), Shirley background (red solid lines) and fitted data after background subtraction (blue solid lines). The XPS spectra was taken after annealing the B: Chromia sample at 550 K. Fig 3-a (a) shows the data from the surface of the sample i.e., d=0 nm. Fig 3-a (b-f) shows the data for (b) d=5.12 nm, (c) d=10.24 nm, (d) d=15.36 nm, (e) d=20.48 nm and (f) d=25.6 nm.*

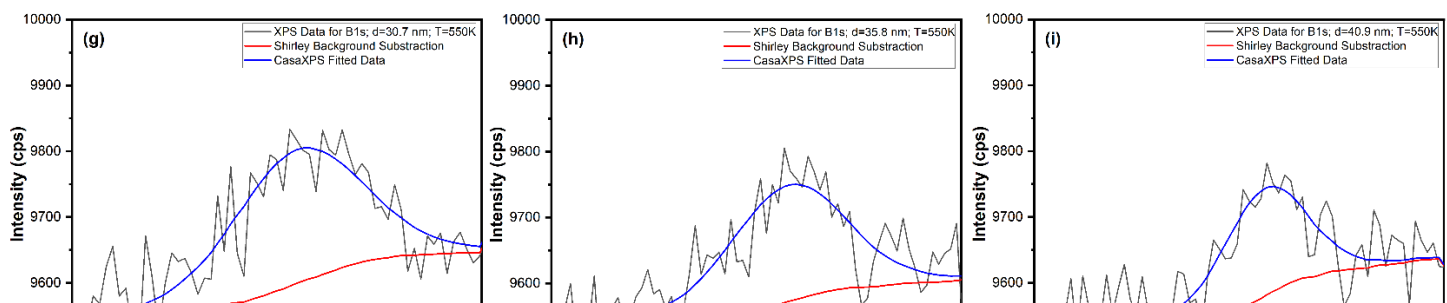

**Figure 7S-b (g-l):** *Figure shows the XPS data for B 1s peaks (black solid lines), Shirley background (red solid lines) and fitted data after background subtraction (blue solid lines). The XPS spectra was taken after annealing the B: Chromia sample at 550 K. Fig 3-b (g) shows the data at the depth of d=30.7 nm, (h) d=35.8 nm, (i) d=40.9 nm, (j) d=46 nm, (k) d=50.12 nm and (l) d= 56.3 nm.*

Similarly **Figure 8S-a (a-f) & Figure 8S-b (g-l)** shows the fitted XPS spectra for Cr $2p_{3/2}$ peaks after background subtraction for the sample annealed at a temperature of 550 K.

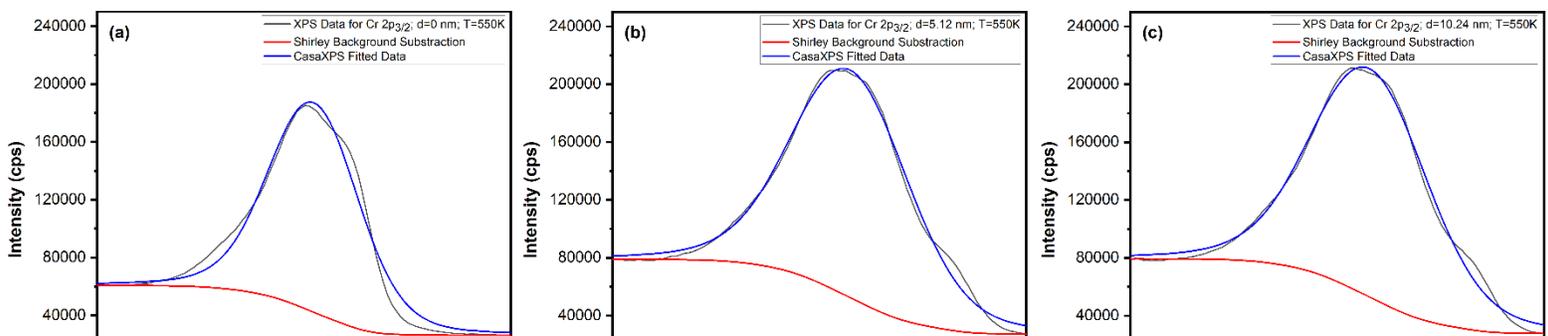

**Figure 8S-a (a-f):** *Figure shows the XPS data for Cr $2p_{3/2}$ peaks (black solid lines), Shirley background (red solid lines) and fitted data after background subtraction (blue solid lines). The XPS spectra was taken after annealing the B: Chromia sample at 550 K. Fig 4-a (a) shows the data from the surface of the sample i.e., d=0 nm. Fig 4-a (b-f) shows the data for (b) d=5.12 nm, (c) d=10.24 nm, (d) d=15.36 nm, (e) d=20.48 nm and (f) d=25.6 nm.*

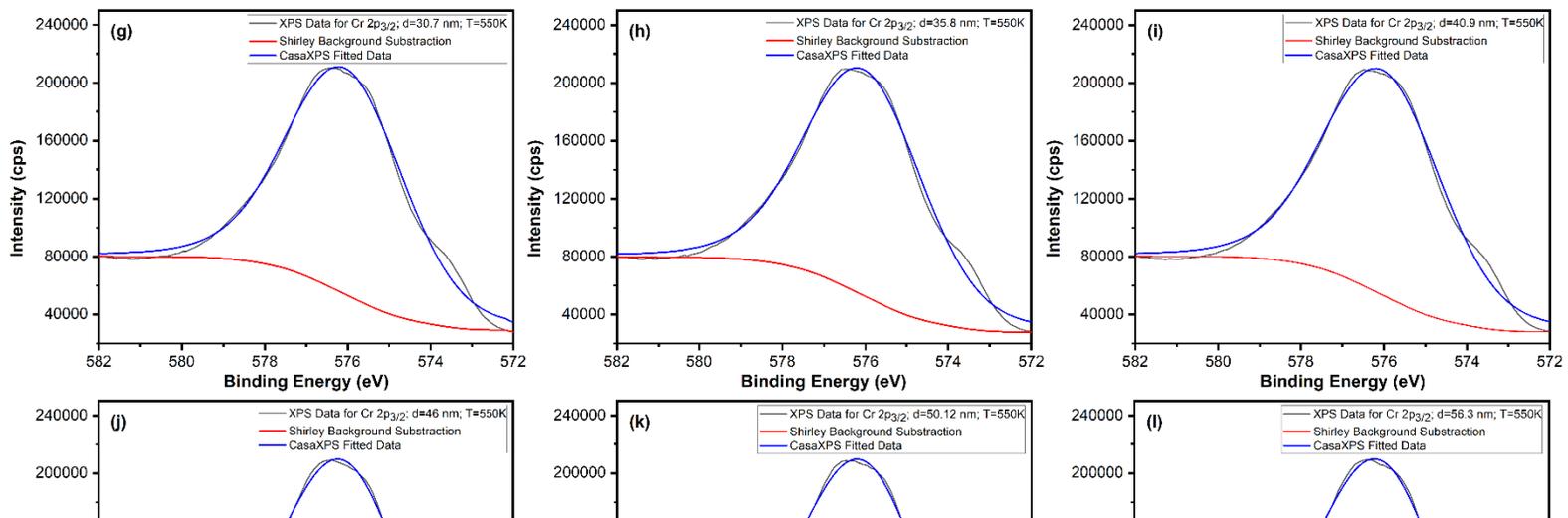

**Figure 8S-b (g-l):** *Figure shows the XPS data for Cr $2p_{3/2}$ peaks (black solid lines), Shirley background (red solid lines) and fitted data after background subtraction (blue solid lines). The XPS spectra was taken after annealing the B: Chromia sample at 550 K. Fig 4-b (g) shows the data at the depth of d=30.7 nm, (h) d=35.8 nm, (i) d=40.9 nm, (j) d=46 nm, (k) d=50.12 nm and (l) d= 56.3 nm.*

Supplementary note 3

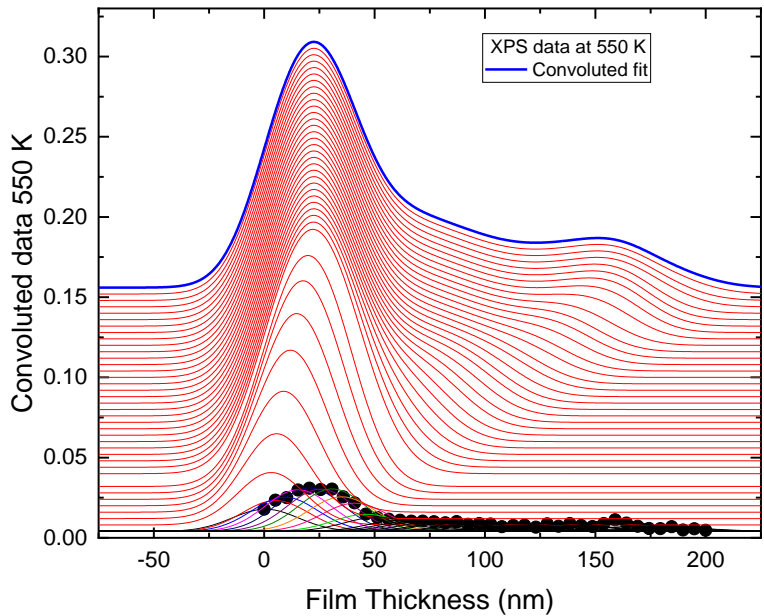

Fig 9S: Convolution technique applied to the XPS data at 550 K from Figure 2a of the main text. Black dots are the original XPS data points. Numerical convolution is done to compare the XPS post annealing data with the cNDP data. To this end, we assign to each XPS data point a Gaussian profile resembling the resolution of the neutron experiment. The height of the respective Gaussian is determined by the y-value of the XPS data point and a FWHM of 30 nm is used to resemble the spatial resolution limitation of the neutron data. The set of Gaussians are added up mimicking a numerical convolution which results in the blue curve. It is in good agreement with cNDP data from figure 1 of the main text. Fig. 6 shows the comparison between the cNDP data and the convoluted XPS data.

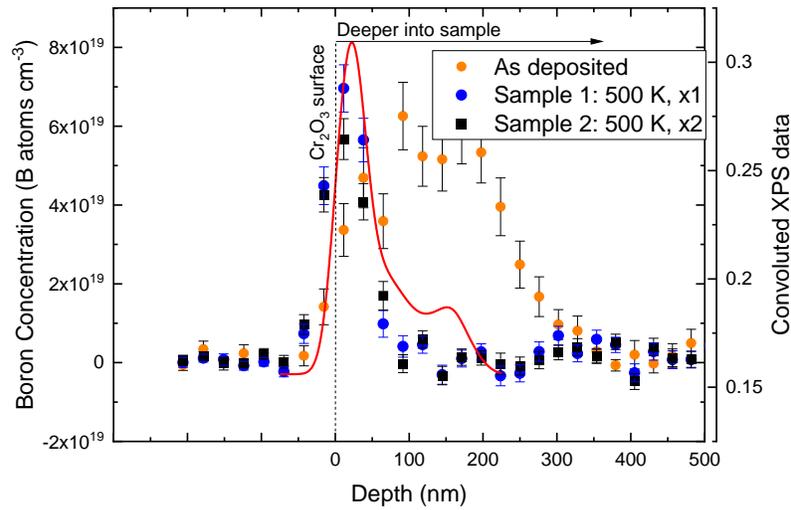

**Fig.10S**: (a) B-concentration depth profile of the as-deposited and heated samples measured via cNDP. Depth = 0 nm is the B:$Cr_2O_3$/vacuum interface and depth ≈ 300 nm is the interface between the as-deposited B:$Cr_2O_3$(0001) film and the $Al_2O_3$(0001) substrate (diamonds). Post-annealing B-concentration depth profiles measured at room temperature after 18 hours (circles) and 36 hours (triangles) annealing at 500 K. Data represented to one standard deviation and are based on experimental counting statistics. The line (right axis) shows the result of a convolution procedure of the XPS depth profiling data. XPS depth profiling measurements are discussed in the subsequent section.

Supplementary note 4

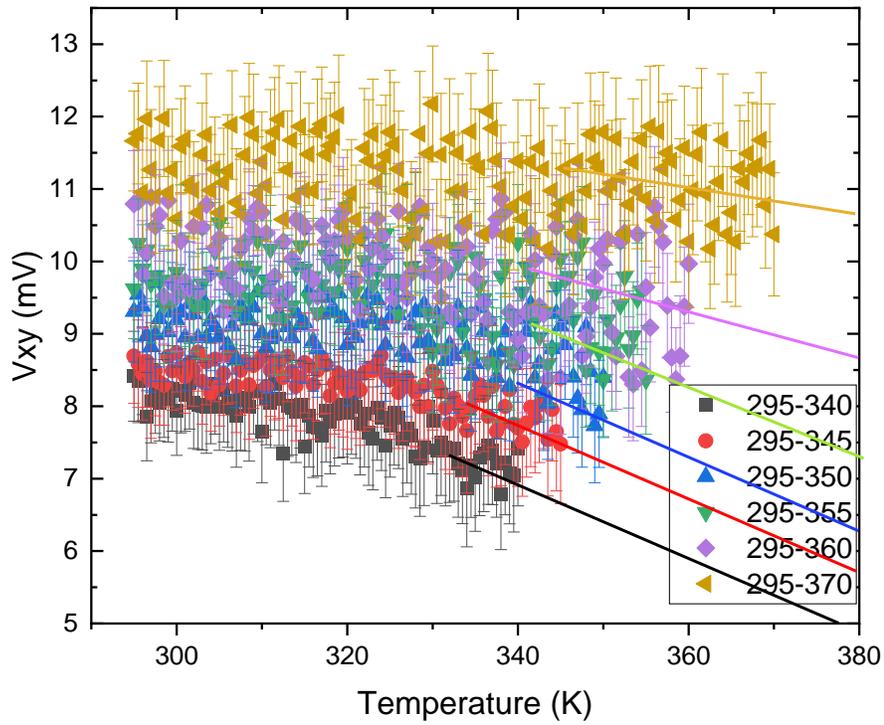

**Fig. 11S**: $V_{xy}$ vs. $T$ for temperature intervals $T_1 = 295K \leq T \leq T_n^{max}$ with $T_1^{max} = 340\ K$ (squares), $T_2^{max} = 345\ K$ (circles), $T_3^{max} = 350\ K$ (triangles), $T_4^{max} = 355\ K$ (down triangles), $T_5^{max} = 360\ K$ (diamonds), $T_6^{max} = 370\ K$ (tilted triangles). Uncertainty bars represent the standard deviation calculated for each set of measurements at each temperature.